\documentclass[9pt]{article}

\usepackage{amsmath,amssymb,amsfonts}
\usepackage{graphicx,color}
\makeatletter

\@addtoreset{equation}{section}
\makeatother

\newcommand{\be}{\begin{equation}}
\newcommand{\ee}{\end{equation}}
\newcommand{\bea}{\begin{eqnarray}}
\newcommand{\eea}{\end{eqnarray}}
\newcommand{\beann}{\begin{eqnarray*}}
\newcommand{\eeann}{\end{eqnarray*}}

\newcommand{\ba}{\begin{array}}
\newcommand{\ea}{\end{array}}

\def\XXint#1#2#3{{\setbox0=\hbox{$#1{#2#3}{\int}$} 
\vcenter{\hbox{$#2#3$}}\kern-.5\wd0}}

\begin{document}

\setlength{\oddsidemargin}{0cm}
\setlength{\baselineskip}{7mm}

\vfil\eject

\setcounter{footnote}{0}

%\tableofcontents
    \begin{Large}  
 \begin{sc}
       \begin{center}
         { Seiberg-Witten prepotential from WZNW conformal block: \\
Langlands duality and Selberg trace formula} 
       \end{center}
    \end{sc} 
        \end{Large}
        
\begin{center}
{ \sc Ta-Sheng Tai}\footnote    
{e-mail address : 
tasheng@riken.jp} 
\end{center}

\begin{small}
\begin{center}
{\it   Theoretical Physics Laboratory, RIKEN,
                    Wako, Saitama 351-0198, JAPAN}
                    
\end{center}
 \end{small}
\begin{abstract}
\noindent
We show how $SU(2)$ $N_f=4$ 
Seiberg-Witten prepotentials are derived form 
$\widehat{\mathfrak{sl}}_{2,k}$ ($k\to2$) four-point conformal blocks via considering Langlands duality.  
\end{abstract}

\section{Introduction}
Last June, Alday, Gaiotto and Tachikawa (AGT) \cite{Alday:2009aq} 
claimed that 
correlation functions of 
primary states in 
Liouville 
field theory (LFT) can 
get re-expressed in terms of Nekrasov's partition function 
$Z_{Nek}$ 
of 4d ${\cal N}=2$ quiver $SU(2)$ SCFT 
(at low-energy Coulomb phase). 
In particular, every 
Riemann surface $C\equiv C_{g,n}$ on which LFT dwells 
is responsible for certain SCFT 
${\cal T}_{g,n} (A_1)$ such that 
the following equality 
\begin{eqnarray*} 
\text{Conformal~block~w.r.t.~}C_{g,n}=\text{Instanton~part~of}~Z_{Nek}\Big({\cal T}_{g,n} (A_1)\Big)
\end{eqnarray*}
holds. 
Their discovery has a profound impact on 
the unification of many known mathematical corners, say, 
Hitchin integrable system (including 
isomonodromic deformation), 
Selberg-Dotsenko-Fateev $\beta$-ensemble and 
$Z_{Nek}$%
\footnote{See \cite{Mironov:2010qe} for explanations between 
the last two topics.}.

For any $C$, talking about the quantization of its moduli space ${\cal M}(C)$ usually relies on a CFT living on it. For 
instance, a finite-dimensional Hilbert space ${\cal H}$ thus obtained is spanned by chiral conformal blocks of, say, WZNW model on $C$ and 
$\text{dim}{\cal H}$ gets computed by Verlinde's formula.  Naively, one wants to ask why only 
CFTs characterized by ${\cal W}$-algebra are more 
preferred than others governed by, say, affine Kac-Moody algebra? 
In this letter, we are going to 
consider 
the role which WZNW models (or conformal blocks of them) play 
in connection with 4d 
${\cal N}=2$ SCFT. It will be shown that at least from 
$\widehat{\mathfrak{sl}}_2$ cases, thanks to their intimate relationship with 
${\cal W}_2/$Virasaro-algebra, the 
$SU(2)$ $N_f=4$ Seiberg-Witten (SW) prepotential 
${\cal F}^{SW}$ is reproduced. It should be 
emphasized that our approach needs neither 
knowledge about AGT conjecture 
nor its ramified (surface operator inserted) version like 
\cite{Alday:2009fs, Kozcaz:2010af, Alday:2010vg, Dimofte:2010tz}. 
What we are truly after is our previous work 
\cite{Tai:2010ps} together with 
a piece of 
independently developed mathematical concept, 
Langlands duality, relating classical LFT and WZNW model at 
critical level. As an intermediate step, 
let us first describe the result of \cite{Tai:2010ps}.

In \cite{Tai:2010ps} we noticed an analogy between 
two seemingly unrelated arenas, say, classical LFT and 
large-$N$ Hermitian matrix model, 
by appealing to Polyakov's conjecture devised for 
$C_{0,4}\equiv  \mathbb{C} \backslash \{ 0,1,q \}$ ($q$: cross-ratio)
\begin{eqnarray} 
&&c(q)=\frac{\partial}{\partial q}
\Big( f_\delta
\Big[
\begin{array}{cc}
\delta_3 & \delta_2 \\
\delta_4 & \delta_1
\end{array}
\Big]
(q) \Big)_{\delta=\delta_s (q)}\nonumber\\
&&f:~s\text{-channel~classical~conformal~block}, 
~~~~~~
{\delta_i}:~\text{classical~conformal~dimension}\nonumber\\
&&\sqrt{{\delta_s}-\frac{1}{4}}=p_s:~s\text{-channel~saddle-point~intermediate~momentum}
\label{pc}
\end{eqnarray}
By translating this formula belonging to LFT 
into the language of 
Hermitian matrix model, 
it had helped us envision%
\footnote{For the sake of brevity, we will simply 
use $f$ which abbreviates the 
classical conformal block.} 
\begin{eqnarray*} 
f\simeq{\cal F}^{SW}
\end{eqnarray*}
The reason is listed as below.\\
${\bf (I)}$~Basically, \eqref{pc} stems from Ward's identity 
involving the $(2,0)$ stress-tensor $T_L (z)=
 {Q}\partial^2_z \phi
-(\partial_{{z}} \phi)^2$ inserted inside 
some four-point Liouville primary field $V_{\alpha}(z)$ 
correlator $\langle X \rangle$ 
under the classical limit $b\to 0$. Besides, 
$\langle X \rangle$ is replaced by 
$\langle 1 \rangle$ when boundary terms ${\cal S}_{bdy}$ fixed by $\Delta_{\alpha}=\alpha (Q-\alpha )$ ($Q=b+{b}^{-1}$) and 
positions of $V_{\alpha}(z)$ are added to 
the original $bulk$ 
Liouville action ${\cal S}_{bul}$. 
Through $b\to 0$, 
$\exp (-{\cal S}_{tot})$ 
(${\cal S}_{tot}={\cal S}_{bul}+{\cal S}_{bdy}$) 
permits one unique saddle-point $\phi_{cl}=\varphi_{cl}/b$ 
whose singular behavior 
is reflected by 
\begin{eqnarray} 
\lim_{b\to 0} b^2 T_L(z)~ {\to} ~T(z)\equiv\frac{1}{2}\partial^2_z \varphi_{cl}
-\frac{1}{4}(\partial_z \varphi_{cl})^2
=\sum_{i=1}^3 \frac{\delta_i}{(z-z_i)^2} + 
\sum_{i=1}^3 \frac{c(z_i)}{z-z_i} 
\end{eqnarray}
where the only unknown 
accessory parameter turns out to be $c(q)$. 
A conjectured 
determination of $c(q)$ is therefore \eqref{pc}. 
This procedure is $rigid$ because of a 
unique $\varphi_{cl}$ which leads to 
a factorized form of 
$\exp (-{\cal S}_{tot}[b^{-1}\varphi_{cl}])=
\exp(-b^{-2}f)\cdot\exp(-b^{-2}\bar{f})\cdots$. 
Only the holomorphic part, $f$, 
survives ${\partial}/{\partial q}$ thereof.\\
${\bf (II)}$~Call the LHS of a 
second-order Fuchsian equation (Baxter equation) 
\begin{eqnarray} 
\partial_z^2 + T(z)=0
\label{1}
\end{eqnarray}
$G$-$oper$. 
In fact, 
\eqref{1} 
is also obtainable 
by imposing 
$b\to 0$ on the null-vector decoupling equation:  
\begin{eqnarray}
\Big( 
L^2_{-1} + b^2L_{-2} \Big) V_{-\frac{b}{2}}=0
\label{nl}
\end{eqnarray}
realized at the conformal block level. 
It seems natural to regard \eqref{1}, 
doubly-sheeted cover of $C_{0,4}$, as 
the genus-zero spectral curve of some Hermitian 
matrix model $Z_M$ through $\partial_z \to iy$ 
due to the same governing Virasoro algebra in both 
situations. 

By doing so, \eqref{pc} necessarily acquires 
certain 
interpretation within the matrix model context. 
To conclude, when \eqref{1} gets identified with 
the large-$N$ spectral curve 
($N$: matrix rank)
\begin{eqnarray*} 
&&y^2=\lim_{N\to \infty} \big\langle T_M (z) \big\rangle 
=\big\langle \partial \phi_{KS} \big\rangle^2\\
&&T_M(z):~\text{stress-tensor~constructed~via~
Kodaira-Spencer~free~boson}~\phi_{KS}(z)~
\text{associated~with}~Z_{M}
\end{eqnarray*}
inevitably $f\simeq {\cal F}_0$, which represents 
genus-zero free energy of $Z_M = \exp(\hbar^{-2} 
{\cal F}_0 + {\cal F}_1 +\cdots)$. 
This sounds plausible because as $\hbar={1}/{N}\to 0$, 
by replacing $f$ with ${\cal F}_0$, 
\eqref{pc} arises from Ward's identity of 
$T_M(z)$ and thus accounts for the accessory parameter of 
the meromorphic function $\lim_{N\to \infty}\big\langle T_M (z) \big\rangle$. 
Moreover, 
\begin{eqnarray}
f\simeq {\cal F}_0={\cal F}^{SW}
\end{eqnarray}
when \eqref{1} $\simeq{\cal G}$ 
(Gaiotto's curve \cite{Gaiotto:2009we}) is assumed 
such that $Z_M$ containing ${\cal G}$ 
(rewritten SW curve) 
as its own spectral curve%
\footnote{See \cite{Dijkgraaf:2009pc, Itoyama:2009sc, Eguchi:2009gf, Fujita:2009gf,Eguchi:2010rf} for recent publications towards 
this direction.} has to be recognized as the instanton part of $Z_{Nek}$ at 
$\epsilon_1 =-\epsilon_2 =\hbar$. In 
\cite{Tai:2010ps} by taking the 
large-$p_s$ limit 
a perfect agreement $f={\cal F}^{SW}$ was observed.

Equipped with these arguments, in the next section 
we start to see 
${\cal F}^{SW}$ indeed also hides 
behind $\widehat{\mathfrak{sl}}_2$ conformal blocks 
via Langlands duality.

\section{Derivation of KZ equation at critical level}
To begin with, what we will mainly rely on is the statement 
encountered in $geometric$ $Langlands$ 
$correspondence$ according to Feigin, Frenkel 
and Reshetikhin \cite{FFR}: \\
$\clubsuit$~``The space of opers, associated with 
Langlands-dual Lie algebra 
${{}^L} {\mathfrak{g}}$ on $\mathbb{P}^1$ with 
regular singularities at marked points $\{ z_i\}$, can be 
identified with the spectrum of a 
corresponding Gaudin 
algebra denoted by ${\cal Z}_{(z_i)}(\mathfrak{g})$.'' \\
Remarkably, the problem of 
fixing accessory parameters of 
a ${{}^L} {G}$-oper (${{}^L}G$: adjoint group of 
${{}^L} {\mathfrak g}$) is switched 
to solving the spectrum of 
${\cal Z}_{(z_i)}(\mathfrak{g})$ (or equivalently 
Gaudin Hamiltonian $\Xi_i$). 
Let us first see how a 
Knizhnik-Zamolodchikov (KZ) equation can be reached 
by means of the proposal $\clubsuit$. 
Ultimately, we are capable of 
claiming that ${\cal N} = 2$ $SU(2)$ $N_f=4$ 
SW prepotentials are encoded in 
$\widehat{\mathfrak{sl}}_2$ four-point conformal blocks 
at critical level $k\to 2$. We stress again that 
any recently developed 
argument from either 
AGT conjecture or its ramified version will not be 
borrowed. Some mathematical aspects will 
be postponed until Sec. 2.2.

Now, given a $PGL_2$-oper 
($PGL_2={}^L {G}$: adjoint group of 
${}^L \mathfrak{g}$ with 
$\mathfrak{g}=\mathfrak{sl}_2$) 
on $C= \mathbb{C}\backslash \{ z_1, \cdots, z_{N}\}$, i.e. 
\begin{eqnarray} 
\partial_z^2 + \sum_{i=1}^N \frac{\delta_i}{(z-z_i)^2} + 
\sum_{i=1}^N \frac{c_i}{z-z_i}
\label{pgl}
\end{eqnarray}
subject to $\sum c_i=0$ 
(no further pole at $z_0=\infty$), 
the pair $(\delta_i , c_i)$ can get read off from 
$(C_i, \Xi_i)$ 
which stands for, respectively, quadratic 
Casimir operators 
\begin{eqnarray} 
C_i=\frac{1}{2}\sum_{a=1}^d J^{(i)}_a J^{a(i)}=
j_i(j_i+1)= -\delta_i=\xi_i(\xi_i-1) 
\label{gh1}
\end{eqnarray}
where $\{ J_a\}$ denotes the 
basis of $\mathfrak{sl}_2$ whilst 
$J^a J_a\equiv \kappa_{ab} J^a J^b$ with $\kappa_{ab}$ being 
the Cartan-Killing form 
and 
Gaudin Hamiltonians 
\begin{eqnarray}
\Xi_i=\sum_{j\ne i}\sum_{a=1}^d
\frac{J^{(i)}_a J^{a(j)}}{z_i - z_j}=c_i
\label{gh2}
\end{eqnarray}
subject to $\sum_i \Xi_i=0$. 
\eqref{gh2} reflects faithfully the above $\clubsuit$.

Notice that 
\eqref{gh1} implies $\xi=j+1$ which had appeared 
in the celebrated $H_3^{+}$-WZNW/Liouville dictionary \cite{} 
near $b\to 0$. Generally, 
the momentum of Liouville primary fields $V_\alpha (z)$ is related to the 
spin-$j$ WZNW primary filed $\Phi^j(y|z)$ by 
\begin{eqnarray*} 
\alpha=\frac{1}{b}(j+1)+\frac{b}{2}.
\end{eqnarray*}
While $y$'s are 
(isospin) variables of the $SL(2, \mathbb{R})$ 
group manifold, a convenient 
spin-$j_r$ representation of 
$\{ J_a^{(r)} \}$ $(a=\pm,3)$ is usually 
chosen like 
\begin{eqnarray*} 
J_{+}^{(r)}=y_r^2\frac{\partial}{\partial y_r}-2j_r y_r, ~~~~~~~~
J_{-}^{(r)}=\frac{\partial}{\partial y_r}, ~~~~~~~~
J_{3}^{(r)}=y_r\frac{\partial}{\partial y_r}-j_r.
\end{eqnarray*}

\subsection{$\widehat{\mathfrak{sl}}_2$ KZ equation}
We are ready to 
show how ${\widehat{\mathfrak{sl}}}_{2,k}$ 
KZ equations on 
a four-punctured ${\mathbb P}^1$ ($q$: cross-ratio)
\begin{eqnarray} 
(k-2)\frac{\partial}{\partial q}\Psi
=\Xi_r \Psi, ~~~~~~~~k-2=\widetilde{b}^{-2} 
, ~~~~~~~~\widetilde{b}=\frac{1}{b}
\label{KZ}
\end{eqnarray}
are obtained. In general, $m$-point spheric $\widehat{\mathfrak{sl}}_{2,k}$ 
conformal blocks satisfy the following KZ equation
\begin{eqnarray} 
&&\Big( 
(k-2)\frac{\partial}{\partial {z_r}}-\Xi_r \Big)
\Psi_m =0,\nonumber\\
&&{\Psi}_m (y|z)\equiv \big\langle \Phi^{j_1}(y_1|z_1)\cdots\Phi^{j_m}(y_m|z_m) \big\rangle.
\label{kkzz}
\end{eqnarray}
At $k\to 2$, we anticipate the solution to 
\eqref{KZ} factorizes as 
\begin{eqnarray*} 
\Psi_4 (y|q) \to  {\mathfrak F}(q)\psi(y)
\end{eqnarray*}
where $\psi(y)$ 
depends only on isospin 
variables and 
\begin{eqnarray*} 
{\mathfrak F}=\exp\big( b^{-2} f\big), ~~~~~~~~
f\equiv f_{\delta}\Big(
\begin{array}{cc}
\delta_3 & \delta_2 \\
\delta_4 & \delta_1
\end{array}
\Big)(q)
\end{eqnarray*}
is referred to as 
the $quantum$ Belavin-Polyakov-Zamolodchikov 
(BPZ) conformal block \cite{BPZ} 
$F_{\Delta}\Big(
\begin{array}{cc}
\Delta_3 & \Delta_2 \\
\Delta_4 & \Delta_1
\end{array}
\Big)$ with $\Delta\equiv b^{-2}\delta$ at $b\to 0$. 
The self-dual symmetry $b \Longleftrightarrow
\widetilde{b}={1}/{b}$ respected by LFT leads to $(k\to 2)\Longleftrightarrow 
(b\to 0)$. While KZ equations are 
analogous to decoupling 
equations of null-vectors at the second level 
(or BPZ systems), a relation 
between \eqref{nl} and \eqref{kkzz} 
(or equivalently duality between 
KZ and BPZ ${\cal D}$-modules over $C$) 
had emerged in 
the framework of quantized geometric Langlands-Drinfeld correspondence. See \cite{Frenkel:2005pa} for an 
excellent lecture note.

Assuming that there exists 
an eigenstate 
$\psi_r (y)$ of $\Xi_r$ and using the 
emphasized relation $c_r=\Xi_r$ in \eqref{gh2}, 
combined with \eqref{pc} one can 
straightforwardly write down%
\footnote{Note that the saddle-point 
intermediate momentum $p_s$ depending on $q$ should be 
substituted after ${\partial}/{\partial q}$ is 
performed.} 
\begin{eqnarray}
  \Big((k-2)\frac{\partial }{\partial q}-
\Xi_r  \Big){\mathfrak F}(q)\psi_r (y)\Big|_{\delta\to \delta_s(q)}=0
, ~~~~~~~~b\to 0.
\label{fff}
\end{eqnarray}
Immediately, this says that 
$\lim_{k\to 2}{\Psi}_4 (y|q)={\mathfrak F}_s(q)\psi_r (y)$ encodes ${\cal F}^{SW}$!

As a matter of fact, at the level of 
$complete$ ($2m$-2)-point spheric correlators 
$\Omega^L$ in LFT as $b\to 0$
\begin{eqnarray}
\Omega^L (z_1,\cdots,z_m|y_1,\cdots,y_{m-2}) 
\to \exp\Big(-{\cal S}_{tot} [b^{-1}\varphi_{cl}] \Big)\prod^{m-2}_{i=1}
\exp\big(-\frac{1}{2}\varphi_{cl}(y_i, \bar{y}_i)\big) 
\label{ccc}
\end{eqnarray}
where again ${\cal S}_{tot}$ receives the boundary 
contribution from $V_{\alpha_i}(z_i)$ insertions 
in addition to ${\cal S}_{bul}$. Besides, 
($m$-2) indicates the total number of 
degenerate $V_{2,1}(z, \bar{z})$ 
evaluated at $\varphi_{cl}$. Furthermore, quoting 
the map of Ribault-Teschner \cite{2} we know 
\begin{eqnarray*}
&&\Omega(z_1,\cdots,z_m)=\Omega^L (z_1,\cdots,z_m|y_1,\cdots,y_{m-2}) ,\nonumber\\
&&\Omega(z_1,\cdots,z_m): 
m\text{-point}~\text{spheric}~\widehat{\mathfrak{sl}}_2\text{-WZNW~correlator}
\end{eqnarray*}
which might, together with \eqref{ccc}, 
play a role in fixing the eigenstate $\psi_r(y)$.

\subsection{Langlands duality}
We have 
seen that $\widehat{\mathfrak{sl}}_{2,k}$ conformal blocks at $k\to 2$ necessarily carries a 
$classical$ piece of LFT through ${\mathfrak F}$ which in turn 
encodes ${\cal F}^{SW}$. 
We want to consequently link two concepts, $G$-oper 
and KZ connection, in order to better understand \eqref{gh2} by resorting to 
a series of arguments below. Namely, this junction joining classical LFT and WZNW model at critical level is truly a piece of Langlands duality in disguise. \\
\\
${\bf (I)}$ Given a Hitchin system defined on $C$ and 
its moduli space ${\cal{M}}_H(C)$, 
to study ${\cal M}_{H}(C)$ amounts to 
examining the so-called Higgs-bundle 
$(d_A^{0,1}, \phi)$ which stems from 
a set of 4d 
$SU(2)$(=${\mathfrak G}$) self-duality equations being 
dimensionally-reduced onto $C$ parameterized by 
$z$ $(\partial^{0,1}_A \equiv 
\bar{\partial}+ A^{0,1})$: 
\begin{equation}
  \begin{cases}
F_A + [\phi, \phi^\ast]=0\\
d^{0,1}_A \phi =0
  \end{cases}  
  \end{equation}
where $F_A$ denotes the curvature of the connection 
$d_A=d+A$. 
The second line implies 
${\mathfrak G}\to 
{\mathfrak G}_{\mathbb C}={\mathfrak sl}(2, \mathbb{C})$ 
such that $\phi$ takes values in Lie algebra of 
${\mathfrak G}_{\mathbb C}$. 
Alternatively, 
${\cal{M}}_H (C)$ is also 
encoded in its spectral curve ($PGL_2$-oper) written as 
\begin{eqnarray} 
\Sigma:~\det \big( y-\phi(z)\big)=0 ~~\to ~~
y^2 + t(z)=0
\label{2}
\end{eqnarray}
with $t(z)$ being a quadratic differential. 
The characteristic polynomial 
\eqref{2} manifests itself as a 
doubly-sheeted cover of $C$; namely, 
each pair $(y,z)$ of $T^\ast C$ is constrained in $\Sigma$. \\
\\
${\bf (II)}$ 
Each Higgs-bundle 
corresponds to 
a flat connection $\nabla$ whose moduli space 
denoted by 
${\cal{M}}_{flat}({\mathfrak G}_{\mathbb C},C)$ 
is actually the space of homomorphisms 
$\pi_1(C) \to {\mathfrak G}_{\mathbb C}$ modulo 
conjugation, i.e. 
\begin{eqnarray*} 
{\cal{M}}_{flat}({\mathfrak G}_{\mathbb C},C) \simeq 
{\cal{M}}_H(C) \simeq {\text{Hom}}(\pi_1 (C), 
{\mathfrak G}_{\mathbb C}).
\end{eqnarray*}
One is capable of choosing a flat connection 
$\nabla$ ($t^a$: ${\mathfrak g}={\mathfrak sl}_2$ 
generator) 
\begin{eqnarray} 
\nabla=\partial_z - \Theta(z), ~~~~~~~~
\Theta(z)=-\sum^N_{i=1}\frac{{\cal A}_i}{z-z_i}, ~~~~~~~~
{\cal A}_i=t^a J^{(i)}_a
\label{ds}
\end{eqnarray}
such that $\det (\nabla)$ gives rise to the RHS of 
\eqref{2} conventionally called Drinfeld-Sokolov form%
\footnote{A properly gauge-transformed 
$\nabla \to C(z) \nabla C(z)^{-1}$ 
leads to the diagonal $\Theta(z)$ called Miura form.}.\\
${\bf (III)}$ As the last step, 
we can construct Schlesinger 
(isomonodromic deformation) equations
from \eqref{ds}, i.e. 
\begin{equation} 
\partial_x \Upsilon =  
\sum^N_{i=1}\dfrac{{\cal A}_i}{x-z_i}\Upsilon , ~~~~~~~~
\partial_{z_i} \Upsilon =  -
\dfrac{{\cal A}_i}{x-z_i}\Upsilon.
\end{equation}
In fact, Schlesinger's system bears another 
equivalent Poisson form:  
\begin{equation} 
\partial_{z_i} {\cal A}_j=\{ \Xi_i, ~{\cal A}_j \}, ~~~~~~~~
\Xi_i=\sum_{j(\ne i)}\dfrac{\text{tr} 
({\cal A}_i {\cal A}_j)}{z_i - z_j}
=\sum_{j(\ne i)}\sum_{a}
\dfrac{J^{(i)}_a J^{a(j)}}{z_i - z_j}
\end{equation}
where the covariant derivative 
$\partial_{z_i} -\{ \Xi_i, ~\cdot \}$ w.r.t. the 
matrix-valued ${\cal A}_i$ is exactly of KZ type. Of course, going from 
\eqref{ds} to its corresponding 
KZ equation can be resorted to 
an inverse procedure of Sklyanin's 
separation of variables \cite{sov}. 
Now the desired link is 
completed. See \cite{Teschner:2010je} 
(and references therein) for very detailed treatment for these subjects.

\section{Discussion: Selberg trace formula vs 
AGT dictionary} 
In \cite{Tai:2010ps} through 
$p_s \equiv a$ 
($a$: $SU(2)$ 
Coulomb parameter) we have 
correctly reproduced ${\cal F}^{SW}$ using the 
$s$-channel classical conformal blocks $f$. This relation 
actually resembles AGT dictionary 
\cite{Alday:2009aq} $classically$. 
The key identification $p_s\equiv a$ says 
equivalently 
\begin{eqnarray} 
\sqrt{\delta_s-\frac{1}{4}}=p_s\equiv \ell_s 
\label{sel}
\end{eqnarray}
due to 
\begin{eqnarray*} 
\ell_s=\oint dz~ \sqrt{T(z)}=a. 
\end{eqnarray*}
Note that $\ell$ satisfying $2\cosh\dfrac{\ell}{2}=\text{tr}
\rho(\gamma)$ 
is the geodesic 
length of some hyperbolic geometry 
${\mathbb H}/\Gamma$ with $\Gamma \subset {\text {Aut}}(\mathbb{H})$ given a homomorphism $\rho:\Gamma=\pi_1(C)\to PSL(2,
{\mathbb R})$ where $\gamma$'s are generators of the 
fundamental group $\pi_1(C)$. Taking 
$C={\mathbb C} \backslash \{ 0,1,q\}$ for example, one has 
a geodesic $\gamma_{12}=\gamma_1 \gamma_2$ encircling two marked points $(z_1,z_2)$ such 
that ($\ell_{12}=\ell(\gamma_{1}\gamma_2)=\ell_s$)
\begin{eqnarray*} 
2\cosh(\frac{\ell_s}{2})=\text{tr}\big( 
\rho(\gamma_1)\rho(\gamma_2) \big).
\end{eqnarray*}

When 
$\delta_s= \xi_s(1-\xi_s)$ is regarded as the eigenvalue 
of the hyperbolic Laplacian $\Delta_C$ defined on 
$C$, \eqref{sel} looks like 
the celebrated Selberg trace formula which 
conjectures that for an 
arbitrary Riemann surface $C={\mathbb H}/\Gamma$ one has  
\begin{eqnarray} 
\sum_{\delta \in \text{Spec}(\Delta_c)}
h(\sqrt{\delta -\frac{1}{4}})=
\sum_{\gamma \in\pi_1(C)}\widetilde{h}(\gamma )
\label{se}
\end{eqnarray}
where $h$ is certain test function and 
$\widetilde{h}$ its suitably-transformed counterpart. 
As \eqref{se} equates 
summations over the 
spectrum of $\Delta_C$ and closed 
geodesic on $C$, it is not actually 
\eqref{sel} until some $good$ test function $h$ 
which reproduces \eqref{sel} is 
selected. 
Nevertheless, 
we believe that \eqref{sel} 
might shed new light on understanding AGT dictionary. 
We hope to return to this issue soon in another 
publication.

\section*{Acknowledgments}
TST thanks each string group 
in Titech, OIQP, Osaka City University and Kyoto University 
for hospitality. Also, TST benefits 
a lot by illuminating communications with Akishi Kato, 
Shigenobu Kurokawa 
and Yasuhiko Yamada.


\begin{thebibliography}{999}
\parskip=-2pt

 \bibitem{Alday:2009aq}
  L.~F.~Alday, D.~Gaiotto and Y.~Tachikawa,
  %``Liouville Correlation Functions from Four-dimensional Gauge Theories,''
  Lett.\ Math.\ Phys.\  {\bf 91} (2010) 167
  [arXiv:0906.3219 [hep-th]].
  %%CITATION = LMPHD,91,167;%%
  
  
\bibitem{Mironov:2010qe}
  A.~Mironov, A.~Morozov and S.~Shakirov,
  %``Towards a proof of AGT conjecture by methods of matrix models,''
  arXiv:1011.5629 [hep-th].
  %%CITATION = ARXIV:1011.5629;%%

\bibitem{Alday:2009fs}
  L.~F.~Alday, D.~Gaiotto, S.~Gukov, Y.~Tachikawa and H.~Verlinde,
  %``Loop and surface operators in N=2 gauge theory and Liouville modular
  %geometry,''
  JHEP {\bf 1001} (2010) 113
  [arXiv:0909.0945 [hep-th]].
  %%CITATION = JHEPA,1001,113;%%

  
  
%\cite{}
\bibitem{Kozcaz:2010af}
  C.~Kozcaz, S.~Pasquetti and N.~Wyllard,
  %``A & B model approaches to surface operators and Toda theories,''
  arXiv:1004.2025 [hep-th].
  %%CITATION = ARXIV:1004.2025;%%


%\cite}
\bibitem{Alday:2010vg}
  L.~F.~Alday and Y.~Tachikawa,
  %``Affine SL(2) conformal blocks from 4d gauge theories,''
  arXiv:1005.4469 [hep-th].
  %%CITATION = ARXIV:1005.4469;%%
%\cite{Dimofte:2010tz}
\bibitem{Dimofte:2010tz}
  T.~Dimofte, S.~Gukov and L.~Hollands,
  %``Vortex Counting and Lagrangian 3-manifolds,''
  arXiv:1006.0977 [hep-th].


  
  \bibitem{Tai:2010ps}
  T.~S.~Tai,
  %``Uniformization, Calogero-Moser/Heun duality and Sutherland/bubbling
  %pants,''
  JHEP {\bf 1010} (2010) 107
  [arXiv:1008.4332 [hep-th]].
  %%CITATION = JHEPA,1010,107;%%
  
  
  %\cite{Gaiotto:2009we}
\bibitem{Gaiotto:2009we}
  D.~Gaiotto,
  %``N=2 dualities,''
  arXiv:0904.2715 [hep-th].
  %%CITATION = ARXIV:0904.2715;%%

  
  \bibitem{Dijkgraaf:2009pc}
  R.~Dijkgraaf and C.~Vafa,
  %``Toda Theories, Matrix Models, Topological Strings, and N=2 Gauge Systems,''
  arXiv:0909.2453 [hep-th].
  %%CITATION = ARXIV:0909.2453;%%

\bibitem{Itoyama:2009sc}
  H.~Itoyama, K.~Maruyoshi and T.~Oota,
  %``The Quiver Matrix Model and 2d-4d Conformal Connection,''
  Prog.\ Theor.\ Phys.\  {\bf 123} (2010) 957
  [arXiv:0911.4244 [hep-th]].
  %%CITATION = PTPKA,123,957;%%

\bibitem{Eguchi:2009gf}
  T.~Eguchi and K.~Maruyoshi,
  %``Penner Type Matrix Model and Seiberg-Witten Theory,''
  JHEP {\bf 1002} (2010) 022
  [arXiv:0911.4797 [hep-th]].
  %%CITATION = JHEPA,1002,022;%%

\bibitem{Fujita:2009gf}
  M.~Fujita, Y.~Hatsuda and T.~S.~Tai,
  %``Genus-one correction to asymptotically free Seiberg-Witten prepotential
  %from Dijkgraaf-Vafa matrix model,''
  JHEP {\bf 1003} (2010) 046
  [arXiv:0912.2988 [hep-th]].
  %%CITATION = JHEPA,1003,046;%%

  \bibitem{Eguchi:2010rf}
  T.~Eguchi and K.~Maruyoshi,
  %``Seiberg-Witten theory, matrix model and AGT relation,''
  JHEP {\bf 1007} (2010) 081
  [arXiv:1006.0828 [hep-th]].
  
 
  
  
  
  
  \bibitem{FFR} B. Feigin, E. Frenkel and 
N. Reshetikhin, ``Gaudin model, Bethe ansatz and critical
level,'' Comm. Math. Phys. {\bf 166} (1994), 27-62. 
 \bibitem{BPZ}
A. A. Belavin, A. M. Polyakov, and A. B. Zamolodchikov, 
``Infinite conformal symmetry in
two-dimensional quantum field theory," 
Nucl.\ Phys.\  B {\bf 241} (1984) 333. 

\bibitem{Frenkel:2005pa}
  E.~Frenkel,
  %``Lectures on the Langlands program and conformal field theory,''
  arXiv:hep-th/0512172.
  %%CITATION = HEP-TH/0512172;%%

  \bibitem{2}
  S.~Ribault and J.~Teschner,
  %``H(3)+ WZNW correlators from Liouville theory,''
  JHEP {\bf 0506}, 014 (2005)
  [arXiv:hep-th/0502048].
  %%CITATION = JHEPA,0506,014;%%

  \bibitem{sov}
    E.K. Sklyanin, ``Separation of Variables 
New Trends, '' [arXiv:solv-int/9504001].  
  
  
%\cite{Teschner:2010je}
\bibitem{Teschner:2010je}
  J.~Teschner,
  %``Quantization of the Hitchin moduli spaces, Liouville theory, and the
  %geometric Langlands correspondence I,''
  arXiv:1005.2846 [hep-th].
  %%CITATION = ARXIV:1005.2846;%%

  
  
  
  
\end{thebibliography}
\end{document}